\documentstyle[11pt,newpasp,twoside,epsf]{article}
\def\edcomment#1{\iffalse\marginpar{\raggedright\sl#1\/}\else\relax\fi}
\reversemarginpar

\begin{document}
\title{XMM-Newton Probes the Solar Past: Coronal Abundances of Solar Analogs at Different Ages}
\author{Manuel G\"udel, Marc Audard, Anton Sres, Rudolf Wehrli}
\affil{Paul Scherrer Institut, W\"urenlingen \& Villigen, CH-5232 Villigen PSI, Switzerland}
\author{Ehud Behar}
\affil{Columbia Astrophysics Laboratory, Columbia University, 550 West 
     120th Street, New York, NY 10027, USA}
\author{Rolf Mewe, Anton J.~J. Raassen}
\affil{Space Research Organization of the Netherlands (SRON),
       Sorbonnelaan 2, 3584 CA Utrecht, The Netherlands}
\author{Hilary Magee}
\affil{Mullard Space Science Laboratory, UCL, Holmbury St. Mary, 
       Dorking, Surrey, RH5 6NT, UK}

\begin{abstract}
We present an X-ray spectral analysis of four solar analogs with different ages and 
magnetic activity levels.  We find largely different coronal compositions. 
The most active stars tend to show an ``Inverse First Ionization Potential'' (IFIP) 
effect, i.e., elements with low FIP are underabundant. Less active stars
tend to the opposite effect, with relative overabundances of low-FIP elements.
Equivalent Chandra results are presented that support these results.
\end{abstract}

\section{Introduction} 

Previous X-ray observations provided evidence for anomalous elemental compositions 
of active stellar coronae when compared to the solar photosphere. Spectra from 
many active stars reveal significant underabundances of most accessible elements (Drake 
1996). This contrasts with the typical solar coronal composition that shows enhancements 
of elements with a First Ionization Potential (FIP) below 10 eV, typically by factors 
of about 4 relative to high-FIP elements and hydrogen (Feldman 1992). 
On  the other hand, spectra from inactive stars revealed either no abundance anomalies, 
or evidence for a solar-like FIP effect (Drake, Laming, \& Widing 1995, 1997), 
suggesting systematic abundance differences in the two classes of stars (Drake 1996). 

New grating spectra, obtained with {\it XMM-Newton}, have added a new riddle to 
coronal abundance studies. Deep observations of HR~1099 (Brinkman et al. 2001, Audard, G\"udel,
\& Mewe 2001), AB Dor (G\"udel et al. 2001a), and YY Gem (G\"udel et al. 2001b) showed   
characteristic underabundances of low-FIP elements such as Fe, Mg, and Si relative to
high-FIP elements such as C, N, and O, with the highest-FIP element Ne showing the 
highest abundance. This pattern,  dubbed the ``Inverse FIP Effect'', challenges 
previous models devised for the element fractionation in the (solar) chromosphere. 
A high Ne/Fe abundance ratio in HR~1099 has been confirmed by {\it Chandra} as well 
(Drake et al. 2001). Since the coronal material ultimately derives from the photosphere, 
a physically interesting result should relate the coronal composition to the photospheric
mixture of elements, which is often unknown. The stars we investigate here, in contrast, are 
known to be of solar photospheric composition. 

\section{Targets and Analysis}

Our two least active stars, $\pi^1$ UMa (G1~V) and 
$\chi^1$ Ori (G0~V), are both members of the Ursa Major Stream, with an 
estimated age of  $\sim$300~Myr (Dorren \& Guinan 1993). Stars in this stream
are of near-solar composition (Soderblom  \& Mayor 1993), supported by 
the measured [Fe/H] values of our targets (Cayrel de Strobel et al. 2001).
EK Dra (dG0e), a Zero-Age Main-Sequence (ZAMS) star with an age of 
$\sim$100~Myr, is a kinematic member of the Local Association, a stellar group 
with solar photospheric  metallicity (Eggen 1983).  
To study activity at its extreme, we added the active ZAMS  star AB Dor 
(Pakull 1981) to our sample, despite its somewhat later spectral type of K0~V. 
It represents saturated activity and is also a member of the Local Association, 
with a measured photospheric metallicity [M/H] $\approx 0.1\pm 0.2$~dex for  
Al, Ca, Fe, and Ni  (Vilhu et al. 1987).

{\it XMM-Newton} (Jansen et al. 2001) observed our targets for
$\approx 36-59$~ksec each. All data were analyzed following standard
procedures within the SAS software.  We principally
used the high-resolution RGS data (den Herder et al. 2001), with a spectral
resolving power of 100$-$500 in the bandpass from 5$-$38~\AA; see Figure 1.
To constrain high-$T$ plasma components and to measure resolved He-like 
and H-like lines of Mg, Si, S, and Ar, we included one of the 
MOS CCD spectra (Turner et al. 2001), but only above 
$\sim$1.4~keV. Since MOS was closed during the AB Dor observation, we used the 
corresponding EPIC PN data instead (Str\"uder et al. 2001). 

\begin{figure}[t!]
\plotone{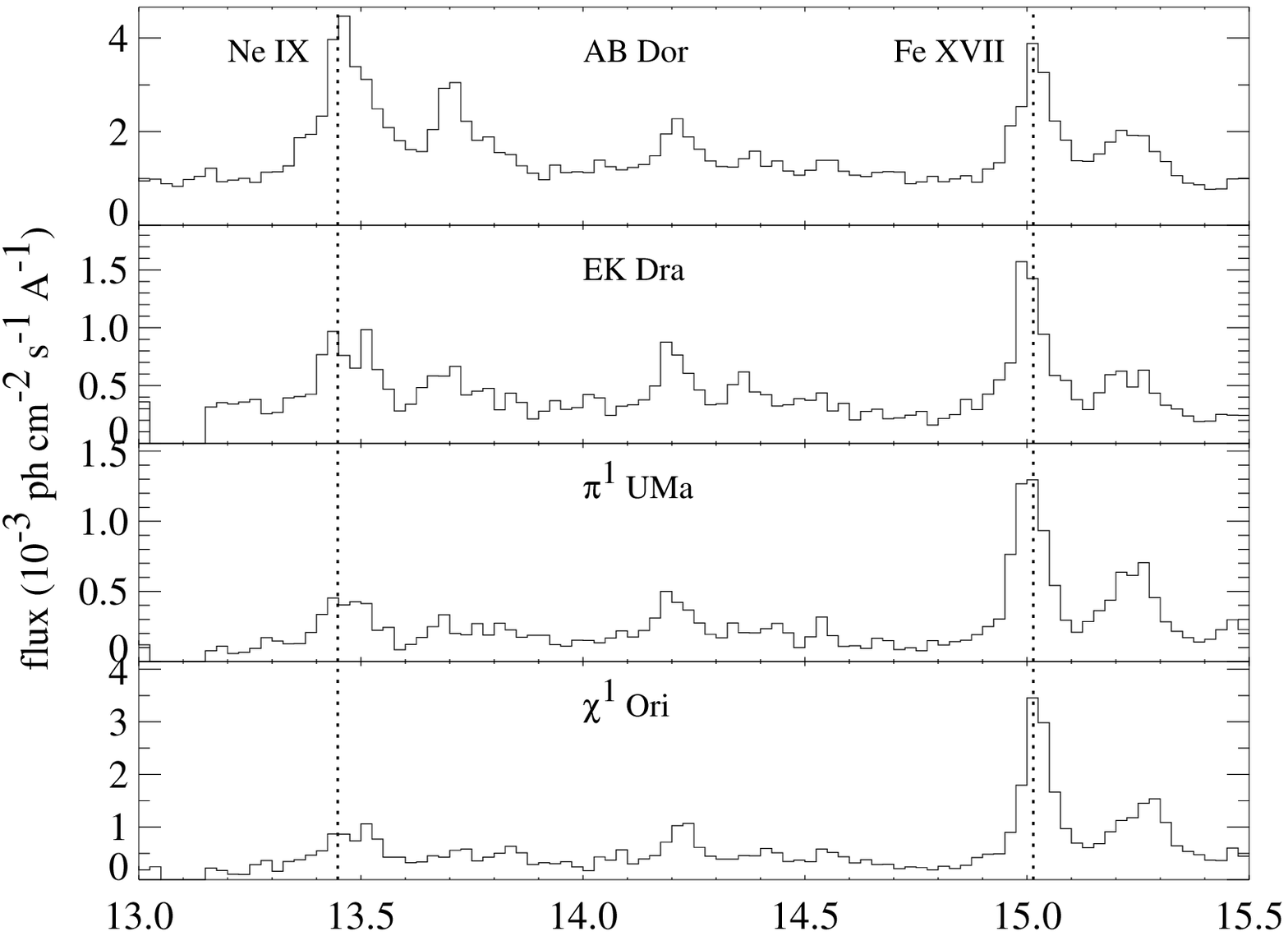}\\
\plotone{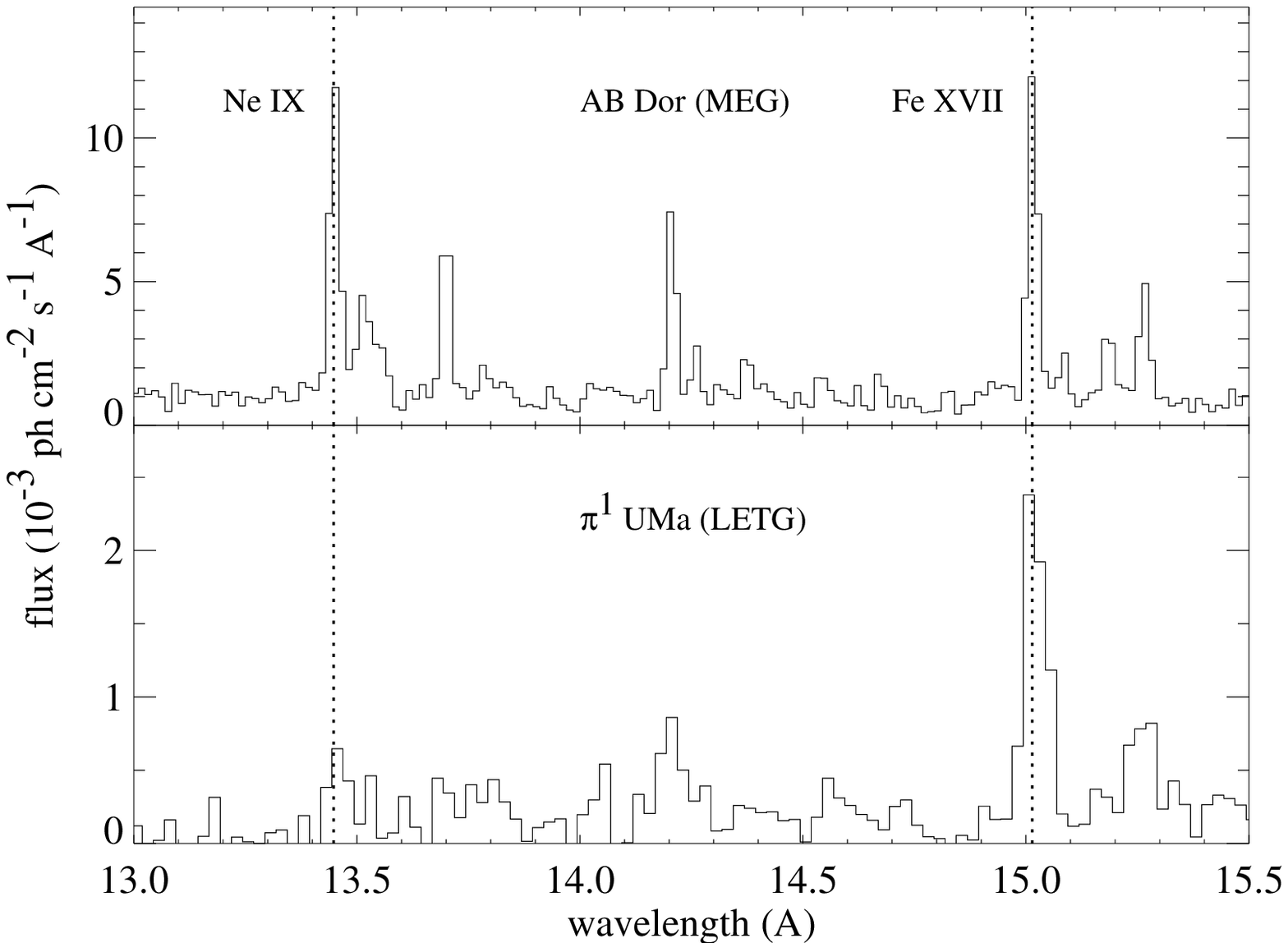}
\caption{Fluxed X-ray spectra between 13$-$15.5\AA\ of solar-type stars; the activity 
 level decreases from top to bottom in each figure. {\bf Top:} {\it XMM-Newton} RGS;
 {\bf Bottom:} {\it Chandra} HETGS/MEG and LETGS. Two lines of Ne~IX and Fe~XVII are marked.}
\end{figure}

\section{Results and Interpretation}

The flux ratios between Ne or O lines and 
the Fe~XVII lines are much smaller in $\pi^1$ UMa and $\chi^1$ Ori
than in AB Dor or EK Dra (Figure 1a). This trend is difficult to explain 
with a broad EM distribution given that the maximum formation temperatures $T_{\mathrm{m}}$
of Ne~IX  and of Ne~X bracket $T_{\mathrm{m}}$ of Fe~XVII, 
{\it unless} the coronae differ in their elemental composition:  
the Fe/O and Fe/Ne ratios must increase toward lower-activity stars. The effect is
not related to time variability or calibration. We show in Figure 1b a similar analysis
for fluxed {\it Chandra} spectra (from the archive, and using standard matrices as provided
by CXC), with similar  line ratios.

A full  spectral analysis with XSPEC (using the vapec code) corroborates our suggestion. 
Figure 2 illustrates elemental abundances relative to O, normalized with the photospheric
abundance ratios (error bars represent formal fit 90\% confidence limits).
There is a clear trend from an inverse FIP effect in AB Dor toward
a normal FIP effect in $\pi^1$~UMa and $\chi^1$ Ori. 
The Fe/O ratio clearly increases with decreasing activity,  by 
about one order of magnitude. However, no trend  is evident for the Ne/O ratio, which resides at
values of $\approx 2-3$.

With the targets used for this study, we are not subject to uncertainty related to stellar 
composition and are led to the conclusion that the overall magnetic 
activity level  alone governs the amount of FIP or IFIP bias. What could induce the
anomalous suppression of low-FIP elements in active stars?
We speculate that high-energy electrons could be important.
Many magnetically active stellar coronae contain a large number of accelerated 
electrons detected by their gyrosynchrotron emission.  While AB Dor is a prolific 
radio source (Lim et al. 1992), EK Dra's radio luminosity $L_{\mathrm{R}}$ is weaker by 
$\sim$one order of magnitude (G\"udel, Guinan, \& Skinner 1997), and for $\pi^1$~UMa,   
$L_{\mathrm{R}} <  0.003L_{\mathrm{R}}$(AB~Dor) (Gaidos et al. 2000). If the electrons 
do not lose all their kinetic energy by radiation in coronal regions, there will be a net 
downward  electron current into the chromosphere. The penetration depth $\ell \propto 
\epsilon^2/n_c$ for electrons of energy $\epsilon$ and a chromospheric layer of constant 
density $n_c$  (Nagai \& Emslie 1984).
For $n_c = 10^{13}$~cm$^{-3}$, we obtain $\ell \approx 10^{17}\epsilon_{\mathrm{keV}}^2/n_c
\mathrm{[cm]} \approx 10-1000$~km for $\epsilon=10-100$~keV, i.e., a significant fraction of the
chromospheric thickness (after equation 13 in Nagai \& Emslie 1984). We require a 
sufficiently small electron flux in order to balance the energy influx in particles
by radiation before heating exceeds $10^5$~K, i.e., to prevent explosive
chromospheric evaporation.
A charge separation is thus built up, together with an electric field that points downward.
Protons and ions are consequently also driven downward, and
the upper layers of the chromosphere from where the coronal material is ultimately
supplied with material, becomes depleted of (typically singly ionized) low-FIP elements
while (mostly neutral) high-FIP elements remain unaffected. 
Transport of the upper chromospheric layers
into the corona by whatever means, e.g., microflares, will thus produce an IFIP biased
coronal plasma.  Only when the electron flux  is large enough to produce 
explosive evaporation of a larger part of the chromosphere will the IFIP enrichment
be quenched, i.e., the composition returns  to near-photospheric, 
in agreement with observations of large stellar flares (e.g., G\"udel et al. 1999).

\begin{figure}[h!]
\plotone{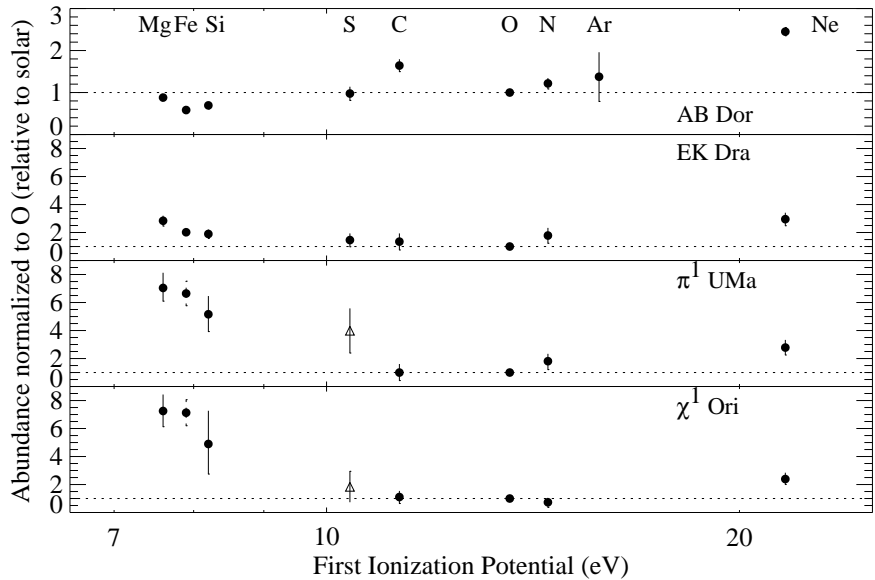}
\caption{Elemental abundances in our four targets, shown as ratios with the   abundance of O
and normalized to the solar photospheric ratios, derived from fits in XSPEC. The
triangles show estimates from individual weak lines (see G\"udel et al. 2002 for further
details).}
\end{figure}

\acknowledgments Research at PSI  has been supported by the Swiss National 
                  Science Foundation (grant 2100-049343). SRON is supported financially by 
		  NWO. The present project is based 
                  on observations obtained with XMM-Newton, an ESA science 
                  mission with instruments and contributions directly funded by 
                  ESA Member States and the USA (NASA).

\end{document}